\documentclass[a4paper]{article}

\usepackage{INTERSPEECH2022}
\usepackage{epstopdf}
\usepackage[utf8]{inputenc}

\usepackage{hyperref}
\usepackage{xstring}
\usepackage{color}
\usepackage{graphicx}
\usepackage{tabularx}
\usepackage{soul}
\usepackage{multibib}
\usepackage{multirow}
\usepackage{float}
\usepackage{blindtext}

\title{Data Augmentation for Low-Resource Quechua ASR Improvement\\}
\name{Rodolfo Zevallos$^1$, Nuria Bel$^1$, Guillermo Cámbara$^1$, Mireia Farrús$^{2,3}$, Jordi Luque$^4$}
\address{
  $^1$Universitat Pompeu Fabra (UPF)\\
  $^2$Centre de Llenguatge i Computació (CLiC), Universitat de Barcelona (UB)\\
  $^3$Institut de Recerca en Sistemes Complexos (UBICS), Universitat de Barcelona (UB)\\
  $^4$Telefónica I+D, Research}
\email{rodolfojoel.zevallos@upf.edu}

\begin{document}

\maketitle
\begin{abstract}

Automatic Speech Recognition (ASR) is a key element in new services that helps users to interact with an automated system. Deep learning methods have made it possible to deploy systems with word error rates below 5\% for ASR of English. However, the use of these methods is only available for languages with hundreds or thousands of hours of audio and their corresponding transcriptions. For the so-called low-resource languages to speed up the availability of resources that can improve the performance of their ASR systems, methods of creating new resources on the basis of existing ones are being investigated. In this paper we describe our data augmentation approach to improve the results of ASR models for low-resource and agglutinative languages. We carry out experiments developing an ASR for Quechua using the wav2letter++ model. We reduced WER by 8.73\% through our approach to the base model. The resulting ASR model obtained 22.75\% WER and was trained with 99 hours of original resources and 99 hours of synthetic data obtained with a combination of text augmentation and synthetic speech generation. 
\end{abstract}

\section{Introduction}

Language Technologies are at the core of the amazing results of Artificial Intelligence recent success. Deep learning technologies have made it possible to use collections of textual and other language resources to create linguistic tools that are widely used in an increasing number of domains. In particular, dialogue systems help users to interact with a system by just speaking their language, which is creating new services in many areas of daily life. For example, conversational agents are being used for medical attention, user’s support in retail, banks and e-commerce applications. 

In recent years, deep learning methods are being applied to the development of Automatic Speech Recognition (ASR) systems with excellent results. However, deep learning methods require large amounts of data, i.e. thousands of hours of fully transcribed audios for each language. Languages that do not count with a sufficient amount of these resources have no options to deploy good enough ASR systems. For these so called low-resource languages, the performance of deep learning models created with a limited amount of data lags far behind that of languages such as English, Spanish, Chinese, etc., with very large amounts of data. Since the collection of data requires of a significant amount of effort and funds, there have been different attempts to explore methods of automatically creating new data on the basis of the available resources to achieve the critical amount of data that can improve the performance of ASR systems. Thus, data augmentation (DA) methods are being developed with some success \cite{Yu-etal2018,Kobayashi2018,wei2019eda,hou-etal-2018-sequence}. 
In this paper we present a DA method to improve the results of ASR models for Southern Quechua. Our method is based on a pragmatic approach combining DA for texts and DA for audios. In DA for texts we generate synthetic texts using the \cite{hou-etal-2018-sequence} DA method but with a variant on the delexicalisation algorithm. In the DA for audios we propose to use the automatically generated texts to generate synthetic audio with a TTS model. Our research aims to contribute to ongoing efforts to support the development of language technologies for indigenous languages of the Americas, most of which are agglutinative and low-resource languages.

The contributions of our research are: (1) a speech corpus for Southern Quechua of 99 transcribed and cleaned hours; (2) an approach combining DA methods for text and DA for audio to generate better diversity in synthetic audios; (3) a variant of the text DA method for resource-poor and agglutinative languages; and (4) an ASR model for Southern Quechua, which to our knowledge is the first deep learning model with acceptable results.

\vspace{-2em}

\section{Quechua Language}
Quechua is a family of languages spoken by about 10 million people in South America, mostly in the Andean regions and also in the valleys and plains connecting the Amazon jungle and the Pacific coast. Quechua languages are considered highly agglutinating with a subject-object-verb (SOV) sentence structure as well as mostly postpositional. 130 suffixes are recognized as common to all dialects. The same suffix may have different forms between dialects, but still fulfill the same function. Their orthography has been normalized among all dialects and the proposal by \cite{CerronPalomino1994} is currently adopted as the Quechua writing standard by the ministries of education both of Peru and Bolivia. 

As for phonetics, Southern Quechua can be mainly divided into two dialects: Chanca Quechua, which has a total of 15 consonants, most of them voiceless, and Collao Quechua, which also has a glottal and an aspirated version of each plosive consonant, leading to a total of 25 consonants. Moreover, the use of voiced consonants present in the phonemic inventory of Spanish is common in all Quechua dialects due to the large number of borrowings. 
\vspace{-1em}
\section{Related Work}

In this section we first review the existing literature for speech systems developed for Quechua, and how they locate concerning the newest advances in the TTS and ASR research areas. Secondly, we introduce the current state of the art in DA. 

\subsection{Text-to-Speech}
Over time in the development of text-to-speech conversion, different techniques have been developed. Today, most TTS models are developed using deep learning models. Tacotron \cite{Wang2017TacotronTE}, a seq2seq architecture for producing magnitude spectrograms from a sequence of characters developed in 2017, simplifies the traditional speech synthesis process by replacing the production of these linguistic and acoustic features with a single neural network trained solely from the training audio. To vocalize the resulting magnitude spectrograms, Tacotron uses the Griffin-Lim algorithm \cite{Tamamori2017SpeakerDependentWV} for phase estimation, followed by a short-time inverse Fourier transform. As the authors point out, this was simply a placeholder for future neural vocoder approaches, as Griffin-Lim produces characteristic artifacts and inferior audio quality compared to approaches such as WaveNet. Tacotron 2 \cite{Shen2018NaturalTS} is an end-to-end model that combines the Tacotron model with a modified WaveNet vocoder \cite{Sotelo2017Char2WavES}, achieving a very natural voice with hours transcribed audio. In this research, as part of the new DA technique, the Tacotron 2 model will be used to create the TTS model for Quechua as it generates a more natural voice and can be used to generate new audios with good quality that can be used for enlarging the ASR training corpus.

\subsection{Automatic Speech Recognition}
Deep learning applied to ASR has achieved state-of-the-art performance in a variety of proposals, based on different building blocks like Transformers \cite{wang2020transformer}, RNNs \cite{Amodei2015, li2019improving} or ResNets \cite{synnaeve2019end}. From these models we highlight the one proposed in the wav2letter++ paper \cite{collobert2016wav2letter}, which is purely convolutional, therefore avoiding the common vanishing/exploding gradients issues from RNNs. Also, the inductive bias in convolutions is helpful in data scarcity scenarios like ours, where Transformer models might underperform due to their need of big data for better generalization \cite{d2021convit}. For such reasons, we use the wav2letter++ model for our ASR baselines.

As for Quechua, \cite{Chacca2018IsolatedAS} developed an ASR model exclusively for numbers using Mel-Frequency Cepstral Coefficients (MFCC), Dynamic Time Warping (DTW) and KNN. It achieved a WER of 8.9\% with a corpus of 5 hours of transcribed audio. The authors in \cite{Cardenas2018SiminchikA} crowdsourced a corpus of 97.5 hours of fully transcribed audio for Southern Quechua and used it for the construction of a pilot ASR system which achieved a 17.2\% WER. This pilot system was developed using the K-Nearest Neighbor (KNN) using the Hidden Markov Model Toolkit \cite{Youngetal2015}. \cite{zevallosetal2020} developed a new ASR model based on the monophone HMM topology, a model with five states per HMM and no jumps, that achieved a WER of 12.7\%. This model was trained using the corpus of 97.5 hours of \cite{Cardenas2018SiminchikA}. Finally, \cite{Aimituma2019} presented the first attempt to train an ASR model for Quechua with deep learning using the Kaldi tool. The model was trained with a corpus of 18 hours of transcribed audio created by themselves, with a WER of 40.8\%.

\subsection{Data Augmentation}

Augmentation methods are techniques that generate synthetic data from an existing dataset. They can generate both text and audio data that would help to cope with the problems that low-resource languages meet when trying to use deep learning methods. The DA techniques to generate synthetic text have been explored mostly to improve text classification systems \cite{Kobayashi2018}, and dialogue and question-answering systems \cite{Yu-etal2018}. \cite{hou-etal-2018-sequence} presented a DA method based on a seq2seq model to generate synthetic text by replacing some words with other words from the same semantic field. Their delexicalisation method replaces segments of a sentence with labels based on their semantic frame. This DA method was successful to improve the accuracy of a language comprehension model for a dialogue system by 6.38\%. \cite{wei2019eda} presented Easy Data Augmentation (EDA) as a set of universal data augmentation techniques. EDA employs four operations (synonym substitution, random insertion, random swapping, and random deletion) to create synthetic texts, improving the experimental classification models by 0.8\% accuracy.

%Data augmentation methods for ASR models are also receiving more and more attention. Among different techniques applied are those that augment the data by modifying the audio features; those that generate new audio using a TTS model; and those that modify the architecture of the ASR model.

Audio distortion methods have been evolving to improve data quality. \cite{Ragni2014DataAF} distorted the vocal tract length (VTLP) managing to improve ASR models by 2.5\% TER (Token Error Rate). \cite{Luetal2020} made a substantial improvement in this technique by distorting noise addition, velocity adjustment and pitch shifting in the original audios, managing to reduce WER by 5.1\%. DA techniques using a TTS model are also delivering good results. \cite{rygaard2015} used a TTS model to generate new audios and improve their ASR for low-resource languages. By synthesizing 4.4 hours on the basis of an English movie subtitles corpus, the improvement of their English ASR was 0.81 absolute points. With 11.3 hours of synthesized blog utterances, the improvement was of 0.93 absolute point WER reduction. \cite{Gokayetal2019} used a technique combining audio distortion and DA to obtain a 14.8\% WER reduction. DA techniques by modifying the ASR model have acquired more and more relevance. \cite{Parketal2019} presented SpecAugment, a DA method that modifies the spectrogram image of the original audio by masking and image warping techniques, achieving a 6\% WER reduction. 
\vspace{-1em}
\section{Methodology}
Our approach consists of three main steps:
\begin{itemize}
  \item{Synthetic text generation: the synthetic text is created using the delexicalisation method proposed by \cite{hou-etal-2018-sequence} but modifying the delexicalisation algorithm so that it can be used for agglutinative and resource-poor languages}.
  \item{Synthetic audio generation: the synthetic text obtained in step 1 is used as input to generate synthetic audio using a TTS model. \cite{Shen2018NaturalTS}}
   \item{Finally, both synthetic text and synthetic audio are aligned and used toghether with the original corpus to train an ASR module with wav2letter++. \cite{Pratap2019Wav2LetterAF}}
 \end{itemize}
\subsection{The Corpora}

For our experiments, we used the Siminchik corpus \cite{Cardenas2018SiminchikA}, with a total of 97.5 hours of fully transcribed audio, and the Lurin corpus\footnote{Both the Siminchik corpus and the Lurin corpus can be found at \url{https://github.com/Llamacha/quechua_resources}}, with a total of 83.3 hours (see Table \ref{tab:lurin}) of fully transcribed audio. The Lurin corpus, has 8,000 sentences. The sentences were collected from different sources: the Collao-Spanish and Chanka-Spanish dictionaries published by the Peruvian Ministry of Education (2021), from Soto's Quechua-English functional dictionary (2007) and from the book Autobiografía de Condori. We used the Quechua morphological parser and normalizer developed by \cite{RiosGonzales2016ABL} to convert the corpus into standard Southern Quechua as in (1). 

\begin{enumerate}
\item[(1)] Original: \textit{Qichwa siminchik kan} \\
Normalized: Qichwa simi-nchik ka-n \\
Literal translation: Quechua mouth-ours is
\end{enumerate}

\begin{table}[H]
\begin{center}
\begin{tabular}{|l|cc|cc|}
\hline
\multirow{2}{*}{}   & \multicolumn{2}{c|}{Female}                                                                                                                & \multicolumn{2}{c|}{Male}                                                                                                                  \\ \cline{2-5} 
                                                             & \multicolumn{1}{c|}{\begin{tabular}[c]{@{}c@{}}Chanka\end{tabular}} & \begin{tabular}[c]{@{}c@{}}Collao\end{tabular} & \multicolumn{1}{c|}{\begin{tabular}[c]{@{}c@{}}Chanka\end{tabular}} & \begin{tabular}[c]{@{}c@{}}Collao\end{tabular} \\ \hline
\begin{tabular}[c]{@{}l@{}}Speakers\end{tabular} & \multicolumn{1}{c|}{24}                                                        & 72                                                        & \multicolumn{1}{c|}{16}                                                        & 48                                                        \\ \hline
\begin{tabular}[c]{@{}l@{}}Hours\end{tabular}   & \multicolumn{1}{c|}{21.3}                                                      & 45.2                                                      & \multicolumn{1}{c|}{7.1}                                                       & 9.7                                                       \\ \hline
\end{tabular}
\caption{Lurin Corpus Statistics.}
\label{tab:lurin}
\end{center}
\end{table}

\vspace{-3em}

Both corpora were preprocessed to improve the quality of the audios, eliminating excess background noise, audios with no voice, those with background music and those not spoken in Quechua, the latter feature added thanks to the first automatic speech recognition developed for Quechua \cite{zevallosetal2020}. Finally, audios longer than 30 seconds were divided into segments of no more than 30 seconds and transformed to mono channel, 16 kHz sampling, 16-bit precision encoding and WAV format. After cleaning the two resources, our corpus for the experiments contained 123.75 hours that were divided as described in the Table \ref{tab:sc}.

\begin{table}[H]
\begin{center}
\begin{tabular}{lcccc}
\hline
\textbf{Corpus}                                             & \textbf{Train} & \textbf{Dev} & \textbf{Test} & \textbf{Total} \\ \hline
\begin{tabular}[c]{@{}l@{}}Siminchik +\\ Lurin\end{tabular} & 150            & 19           & 19            & 188            \\ \hline
Preprocessed                                                & 99             & 13           & 12            & 124            \\ \hline
\end{tabular}
\caption{Speech Corpus Statistics.}
\label{tab:sc}
\end{center}
\end{table}

\vspace{-4,5em}

\subsection{Synthetic text generation}
\label{sub:STG}

We describe text DA methodology for agglutinative languages based on the study of DA for language understanding (LU) using semantic frames \cite{hou-etal-2018-sequence}. The methodology is composed of 4 methods: delexicalisation, diversity classification, text generation using a seq2seq model and surface realization.

In our experiments, we used the corpus shown in the table \ref{tab:sc} with a total of 8,320 sentences, and a Quechua lexicon \cite{Rudnick2011TowardsCW} that provides us with the labels of the semantic frames. The sentences were preprocessed using a tokenizer, a POS tagger and a lemmatizer for Quechua \cite{RiosGonzales2016ABL} in order to find the lemmas to be tagged with their corresponding semantic frame. Differently to \cite{hou-etal-2018-sequence}, in case of not finding the semantic frame of the lemmatized word, we used a Quechua-English bilingual dictionary to find the corresponding one for the Quechua lemmatized word in English. This English word is labeled with its semantic frame using the NLTK library. After obtaining the lemmas with their semantic frames, we evaluate the most frequent semantic frames in all sentences. The most frequent semantic frames are labeled following the BIO standard \cite{ramshaw1999text}. In our experiment, we found that most of the sentences had the following tags: "month-name", "city-name" and "time-name". An example of the list of tagged words is shown in Table \ref{tab:dele}.

The words that were tagged using the BIO standard are added to a list. This list will later be used in the process of filling in the delexicalised sentences generated by the seq2seq model.

After delexicalisation, we used a seq2seq model to generate new delexicalised sentences. Like paraphrase methods, the purpose of using a seq2seq model is to generate phrases that are relatively different from the input phrase. For example, by input d (delexicalized phrase) the seq2seq models generate an output d' (delexicalized phrase modified by paraphrase). We used OpenNMT \cite{Klein2017OpenNMTOT} as the implementation of the seq2seq. We followed the same steps as \cite{Luong2015EffectiveAT} to train the seq2seq model. The number of layers for the LSTM was set to 2 and the size of hidden states to 1000. The dropout of the seq2seq model was set to {0, 0.1, and 0.2} considering its regularization power on small data sizes. The batch size was set to 16; furthermore, the GridsearchCV method of the Sklearn library was used to find the best set of hyperparameters. The algorithm Adam was used as the optimization algorithm of loss.

After training the seq2seq model, we use it to generate new delexicalised sentences. These sentences generated by seq2seq are added to a list. The sentences in the list are ordered using the similarity algorithm of NLTK library. Only the sentences with the least similarity will be used since our objective is to generate phrases with the same semantic framework but different from the original sentence. We evaluate our data augmentation with F-score just like Hou et al. \cite{hou-etal-2018-sequence}. Finally, the tags of the new sentence are replaced by a word with the same tag from the list of tagged words in the delexicalisation process. We generated 8,320 new sentences using the steps mentioned above. 

\begin{table}[H]
\begin{center}
\begin{tabular}{|l|l|l|l}
\cline{1-3}
\multicolumn{1}{|c|}{\textbf{Quechua Word}} & \multicolumn{1}{c|}{\textbf{Semantic Frame Code}} & \multicolumn{1}{c|}{\textbf{English Word}} &  \\ \cline{1-3}
Qayna                                       & I-date month-name                                 & Last                                       &  \\ \cline{1-3}
p'unchay                                    & B-date month-name                                 & day                                        &  \\ \cline{1-3}
Qusqupim                                    & B-from loc city-name                              & in Cuzco                                   &  \\ \cline{1-3}
wawqiykunata                                & O                                                 & to my brothers                             &  \\ \cline{1-3}
watukurqani                                 & O                                                 & I visited you                              &  \\ \cline{1-3}
kunan                                       & I-date month-name                                 & this                                       &  \\ \cline{1-3}
p'unchaytaq                                 & B-date month-name                                 & day                                        &  \\ \cline{1-3}
Punomanmi                                   & B-from loc city-name                              & to Puno                                    &  \\ \cline{1-3}
risaq                                       & O                                                 & I will go to                               &  \\ \cline{1-3}
\end{tabular}
\caption{Example of a delexicalised Quechua sentence; also, its translation into English.}
\label{tab:dele}
\end{center}
\end{table}

\vspace{-4,5em}

\subsection{Synthetic audio generation}

We trained a TTS model for Quechua based on the Tacotron 2 model \cite{Shen2018NaturalTS} in order to achieve good performance with 15 hours of transcribed audio recorded by a single speaker from the corpus used in this research (see section 4.1). The model was trained in 1500 epochs with 1 GPU, the learning rate was 0.001, a batch size of 32 was used, the Adam optimizer algorithm was also used and finally we applied a gradient clipping threshold of 0.1. The Quechua TTS model achieved a mean opinion score (MOS) of 3.15\% which was obtained by consulting Quechua speakers from southern Peru. We used the Quechua TTS to generate 99 hours of synthetic audio from the 8,320 synthetic sentences (described in section \ref{sub:STG}).

\subsection{Training of Southern Quechua ASR system}

The Quechua ASR model was trained with the wav2letter++ toolkit \cite{Pratap2019Wav2LetterAF} for 242 epochs. Stochastic gradient descent with Nesterov momentum was used along with a minibatch of 4 utterances. The initial learning rate was set to 0.002 for faster convergence and it was annealed with a constant factor of 1.2 after each epoch, with momentum set to 0. The architecture of the ASR model is the one from the Conv GLU model proposed in the wav2letter++ WSJ recipe \cite{collobert2016wav2letter}. The model was optimized following the Auto Segmentation Criterion (ASG) \cite{collobert2016wav2letter}.

The created vocabulary contains 34 graphemes: the standard Spanish alphabet plus the apostrophe, silence and 6 graphemes of vowels with accents and diaeresis, plus a lexicon with all the words of the corpus separated at letter level. The hyperparameters of the architecture, as well as those of the decoder, were adjusted using the validation set. The MFCC features were computed with 13 coefficients, a 25 ms sliding window and a 10 ms interval. First and second order derivatives were included. Power spectrum features were calculated with a 25 ms window, 10 ms interval and 257 components. All features were normalized (mean 0, std 1) per input sequence.

The Southern Quechua ASR model for our experiment was trained with 99 hours of original transcribed audio and 99 hours of the synthetic transcribed audio obtained from the 8,320 synthetic sentences. The 8,320 synthetic sentences were added to the Siminchik corpus, as used by \cite{Cardenas2018SiminchikA}, to train the ASR language model using the modified Kneser-Ney n-gram model \cite{Heafield2013ScalableMK}. Given the high number of hapax, a singleton pruning rate with a "K" of 0.04 was used to randomly replace only a "K" fraction of the once-occurring words in the training data with a global UNK symbol. The word-level 4-gram language model developed showed a perplexity of 282.45. 

\subsection{Baselines}

\subsubsection{Original data Baseline}
\vspace{-0.5em}
An ASR model with the same parameters as described in the previous section but trained with only the original 99 hours of transcribed audio of our corpus as shown in table 4 was used as a baseline to see WER improvement.
\vspace{-0.5em}
\subsubsection{Distortion Baseline}
\vspace{-0.5em}
As a second baseline, the method of DA through audio distortion proposed by \cite{Luetal2020} was developed to augment the speech data of the original corpus. The nlpaug13 library was used to manipulate the training audios (99 hours) by modifying the speed according to a randomly selected coefficient in the range between 0.85 and 1.15, which is where this DA technique performs best. By applying this speed distortion to the training audios, we were able to duplicate the training data to 198 hours of transcribed audio. The ASR model of Quechua was trained with the same configuration as explained in section 4.4.
\vspace{-0.5em}
\subsubsection{More Data Baseline}
\vspace{-0.5em}
As a third baseline, the ASR model was trained following the same steps of the Original Baseline but repeating the 99 hours of transcribed audio, i.e., 198 total hours of original transcribed audio but with repetitions. The language model of \cite{Cardenas2018SiminchikA} was used for this baseline. 

\section{Results}
Four experiments were conducted to evaluate the new DA method using a seq2seq model to generate synthetic text and a TTS model to generate synthetic speech, in order to improve the result of the ASR model of Southern Quechua as trained with the original data available. The results of the experiments are shown in Table \ref{tab:result}.

\begin{table}[htbp]
\begin{center}
\begin{tabular}{lcc}
\hline
\textbf{Data}  & \textbf{Training Hours} & \textbf{WER (\%)} \\
\hline
Original Data  & 99                      & 31.5                \\
Distorted Data & 99+99 (distorted)       & 25.1                \\
More Data      & 99+99 (original)        & 26.1                \\
Synthetic Data & 99+99 (synthetic)       & 22.8     \\ \hline          
\end{tabular}
\caption{Results of the experiments.}
\label{tab:result}
\end{center}
\end{table}
\vspace{-3em}

The baseline ASR model used 99 hours of transcribed audio as described in section 4.4 achieved a WER of 31.5\%. The distortion baseline managed to decrease the WER from 31.5\% to 25.1\%, that is a relative improvement of 6.4\% in WER. The more data baseline achieved a  reduction of WER from 31.5\% to 26.1\%. Therefore, a relative improvement of 5.4\% was obtained. However, better results are achieved with the distorted audio baseline. The results of our DA method showed a decrease of the WER of the ASR model from 31.5\% to 22.8\%. This represents a relative improvement of 8.7\% validating that it is the most successful approach.

\section{Analysis of the results}

Our DA method was compared with other DA methods to verify whether it could obtain better results for the ASR model of Quechua. Our DA method achieved the best results with a relative improvement of 8.7\% WER reduction with respect to the original dataset baseline. 

Our DA method generates synthetic text and audio that contributes to improve the ASR model of Quechua. With the More Data Baseline experiment we prove that it is not just an effect of having more data, because by duplicating the original data to train the ASR model, the improvement is lower than training the model with synthetic data. Finally, our results also show that synthetic data is a more effective approach than the distortion method that achieved lower WER reduction. 

In order to better understand the results of our DA method, we performed two more experiments. We were curious to see the role of the new language model, also trained with synthetic data, in the results of the ASR system. For that purpose, first, we run an experiment with the ASR model trained only with the original data and with the new Language Model, that is the one trained with also synthetic texts; second, we run another experiment with the ASR model trained with the original and the synthetic data and the old language model, i.e. without synthetic data. 
The experiment one got 29.7\% WER results. Experiment two got 25.6\%, proving that the major contribution comes from the synthetic audio created from the seq2seq generated texts. 

It is important to mention that we have performed the experiments using the WER metric in order to compare our contribution with other models. In future work we will use the TER (Token error rate) metric, because it evaluates a subword (token) within a word and not the whole word. This is more convenient for agglutinative languages since words are usually composed of subwords (suffixes).
\vspace{-1em}
\section{Conclusions}
In this research, a DA method was developed to improve the results of ASR models for agglutinative and low-resource languages such as Quechua. A delexicalisation method for agglutinative languages was built as part of the DA method; furthermore, a language model was developed with the new synthetic texts generated by the DA seq2seq model; also, a TTS based on the Tacotron 2 model was built to generate synthetic audio and finally an ASR based on the wav2letter model was developed.
Some experiments were performed with the original training data and with the DA using the different methods in order to compare the performance of the new DA method.
The results showed that the DA technique works well for improving speech recognition systems in the case of agglutinative and low-resource languages. In this case, a relative improvement of 8.73\% of WER was obtained using the combination of text DA and synthetic speech DA.
\vspace{-1em}
\section{Acknowledgements}

We thank all the Quechua speakers and in particular to Alex Peiró who collaborated in this research. This work has been partially supported by the Project PID2019-104512GB-I00, Ministerio de Ciencia e Innovación and Agencia Estatal de Investigación (Spain), and the INGENIOUS project from the European Union’s Horizon 2020 Research and Innovation Program under grant numbers 833435. The third author has been funded by the Agencia Estatal de Investigación (AEI), Ministerio de Ciencia, Innovación y Universidades and the Fondo Social Europeo (FSE) under grant RYC-2015-17239 (AEI/FSE, UE).

%\newpage

\bibliographystyle{IEEEtran}

\bibliography{mybib}

\end{document}